\documentclass[twocolumn, dvipsnames]{article}

\AtBeginDocument{%
  }

\usepackage{hyperref}
\usepackage{amsmath,amsfonts}
\usepackage{algorithmic}
\usepackage{graphicx}
\usepackage{textcomp}
\usepackage{listings}
\usepackage{hyperref}
\usepackage[utf8]{inputenc}
\usepackage{xcolor}
\usepackage{authblk}

\definecolor{codegreen}{rgb}{0,0.6,0}
\definecolor{codegray}{rgb}{0.5,0.5,0.5}
\definecolor{codepurple}{rgb}{0.58,0,0.82}
\definecolor{backcolour}{rgb}{0.95,0.95,0.92}

\lstdefinestyle{pocstyle}{
    backgroundcolor=\color{backcolour},   
    commentstyle=\color{codegreen},
    keywordstyle=\color{magenta},
    numberstyle=\tiny\color{codegray},
    stringstyle=\color{codepurple},
    basicstyle=\ttfamily\footnotesize,
    breakatwhitespace=false,         
    breaklines=true,                 
    captionpos=b,                    
    keepspaces=true,                 
    numbers=left,                    
    numbersep=5pt,                  
    showspaces=false,                
    showstringspaces=false,
    showtabs=false,                  
    tabsize=2,
    frame=tb
}

\graphicspath{{./img/}}

\begin{document}

\title{Lessons learned from a Performance Analysis and Optimization of a multiscale cellular simulation}


\author[1]{Marc Clascà}
\author[1]{Marta Garcia-Gasulla}
\author[1]{Arnau Montagud}
\author[1]{Jose Carbonell Caballero}
\author[1,2]{Alfonso Valencia}

\affil[1]{Barcelona Supercomputing Center (BSC)}
\affil[2]{ICREA, Pg. Lluís Companys 23, Barcelona}
\affil[ ]{\textit{\{marc.clasca,marta.garcia,arnau.montagud,jose.carbonell,alfonso.valencia\}@bsc.es}}

\date{}


\maketitle

\makeatletter
\def\blfootnote{\gdef\@thefnmark{}\@footnotetext}
\makeatother

\blfootnote{© Marc Clascà et. al. | ACM 2023. This is the author's version of
the work. It is posted here for your personal use. Not for redistribution. The
definitive Version of Record was published in Proceedings of the Platform for
Advanced Scientific Computing Conference, https://doi.org/10.1145/3592979.3593403.}

\section*{Abstract}
This work presents a comprehensive performance analysis and optimization of a multiscale agent-based cellular simulation. 
The optimizations applied are guided by detailed performance analysis and include memory management, load balance, and a locality-aware parallelization. 
The outcome of this paper is not only the speedup of 2.4x achieved by the optimized version with respect to the original PhysiCell code, 
but also the lessons learned and best practices when developing parallel HPC codes to obtain efficient and highly performant applications, especially in the computational biology field.

\section{Introduction}
Multiscale modelling has emerged as one of the most effective approaches for simulating physical systems that comprise different time and spatial scales. 
For this purpose, the interplay between the different levels is carefully modelled, thus determining how particular events at a specific level affect events from other scales.
In computational biology, multiscale modelling has 
been useful for modelling parts of tissues, including physical barriers (such as blood vessels) and different types of cells that communicate both by direct contact and through the production and consumption of different chemical molecules across the extracellular space \cite{Metzcar_review_2019}.

One of the most popular multiscale modelling frameworks is agent-based modelling (ABM) that defines a set of rules that govern the behaviour of each individual element (or agent) in the simulation \cite{Macal_North_2010}.
ABM explicitly defines the interactions among agents and with the environment, and these interactions can produce emergent properties that reproduce the well-known properties that characterize the real system being modelled \cite{Chaste_2020,physicell_2018}.
In addition, ABMs have also been embedded with other models to simulate a cell's signalling pathways, for instance using Boolean modelling \cite{PB2_2022,Cognac_2015}, adding another scale to the multiscale framework.

ABMs' setup have allowed them to be widely used to model cancer growth and drug treatments and its flexiblity has allowed researchers to model bigger, more complex problems that have increased computational requirements.
Indeed, to be able to simulate real-sized tumours with complex micro-environment, there is a need for ABMs models that scale well in computing clusters \cite{Montagud_gigascale_2021}.
In spite of some efforts, most ABM tools were designed for desktop computers and have a considerable room for improvement 
if used in computing clusters \cite{Saxena_biofvmx_2021,Cytowski_2017,FLAME_2012}.
Thus, if we aim at scaling these ABMs efficiently in computing clusters, we first need to diagnose the code and analyze its performance.

Detailed performance analysis is a necessary process in the optimization cycle. Timing and profiling are common approaches used to determine performance; however, they do not provide enough information and are far from revealing insights into the real bottlenecks, pitfalls and sources of inefficiency~\cite{scalasca}.
On the other hand, detailed performance analysis based on execution traces~\cite{performance_opt} can unveil inefficiencies coming from different sources: the parallel code, the hardware, or the system software (f.i., parallel programming models, parallel libraries)~\cite{pdp_efficiency_metrics}. 

In this work, we follow a performance analysis methodology based on a set of efficiency metrics. These metrics indicate the primary source of efficiency loss. Based on this, the analyst can study in detail the execution to point out the main issue and offer suggestions to overcome it. Once the bottlenecks have been identified, we provide the corresponding optimizations and evaluate them~\cite{dmrg}.

The performance issues analyzed and optimized include memory management, load balance, and data locality. These optimizations lead us to a 2.4x speedup with respect to the original version of the code when using a full node of Marenostrum4 (48 cores).
However, the outcome of this work goes beyond the achieved speedup, as it has allowed us to identify common patterns that lead to performance degradation in highly parallel codes. We present the conclusions of our study as a set of best practices that will help scientists and code developers achieve efficient codes.

\section{Related Work}

ABM have been quite ubiquitous in computational biology~\cite{castiglione2014modeling} and there are many tools available to simulate cells using center-based models (or overlapping spheres) and with a surrounding environment simulated explicitly. For a comprehensive review, refer to the work by \cite{Metzcar_review_2019}.

PhysiCell \cite{physicell_2018} is an open-source, flexible and lattice-free agent-based tool for the multiscale simulation of multicellular systems that stands out for lightweight, very efficient and self-contained framework. 
In addition, researchers can build add-ons for PhysiCell, such as PhysiBoSS \cite{physiboss_2019, PB2_2022}, that allows the embedding of Boolean models as gene regulatory networks into each agent.
PhysiCell has been widely used in computational biology to study cancer growth and how immunogenicity enables tumour cells at the outskirts to escape immune attack \cite{Ozik_EMEWS_2018}, to discover how dynamic regimes can counter the tumour cells’ resistance to tumour necrosis factor (TNF) \cite{physiboss_2019} or to optimize these dynamic drug treatments \cite{PdL_EMEWS_2021, Akasiadis_EMEWS_2021}.

Chaste \cite{Chaste_2020} is an open-source, general-purpose simulation package for modelling soft tissues and discrete cell populations.
This tool allows using different modelling frameworks on a given problem, enabling users to select the most appropriate one for their research and to better understand the limitations of each one of them. 
Chaste can also be expanded, for instance, to simulate gene regulatory networks \cite{Cognac_2015} and has been used for different projects, such as intestinal \cite{Dunn_2013} or colonic crypt \cite{Dunn_2012} studies.

FLAME \cite{FLAME_2012} is an open-source, generic framework for agent-based modelling by using finite-state automata with memory. 
This tool has been adapted to be used with distributed GPUs using the OpenCL standard \cite{Richmond_2010}. 
Examples of uses of FLAME range from immunogenic studies \cite{Kabiri_2019} to epidermis modelling \cite{Li_2013}.

Timothy \cite{Cytowski_2014, Cytowski_2015} is an open-source tool able to perform large simulations in cancer projects and models with nuclear-cytoplasmic oscillations of NF-$\kappa$B \cite{Szymanska_2018}.

BioDynaMo \cite{Breitwieser_2021} is an open-source simulation tool able to offload computation to hardware accelerators and load balance agents and their environment. Its extensible and modular desgin allows it to be used in very different fields such as neurite growth, tumour growth and epidemiology.

Biocellion \cite{Kang_2014} is a flexible agent-based simulation framework that has been used to model a wide range of multicellular biological models, such as a bacterial system in soil aggregates and cell sorting simulations. 
Biocellion is freely available for academic use and its single-node version is open source.

Even though a comprehensive, community-driven benchmark of tools is an endeavour still to be addressed in the field, most of these tools compare their characteristics with the rest of the tools in descriptive tables. For instance, Chaste shows this in Table 1 \cite{Chaste_2020} and PhysiCell has a similar table in its Supplementary material \cite{physicell_2018}.
Some tools go a step beyond and present their technical performance under different biological setups, such as FLAME \cite{FLAME_2012}, or computer architectures, such as Timothy \cite{Cytowski_2017}.
In spite of these efforts, a detailed performance analysis aimed at guiding the optimization of any of these tools, as we present in this work, has not been addressed yet.

\section{Background}
\subsection{Computer cluster setup}
MareNostrum4 is a supercomputer based on Intel Xeon Platinum processors from the Skylake generation. 
It is a Lenovo system composed of SD530 Compute Racks, an Intel Omni-Path high performance network interconnect and running SuSE Linux Enterprise Server as operating system. 
Its current Linpack Rmax Performance is 6.2272 Petaflops.

The general-purpose block that we used for present work consists of 48 racks housing 3456 nodes with a grand total of 165,888 processor cores and 390 Terabytes of main memory. 
Compute nodes are equipped with

\begin{itemize}
    \item 2 sockets Intel Xeon Platinum 8160 CPU with 24 cores each @ 2.10GHz for a total of 48 cores per node
    \item L1d 32K; L1i cache 32K; L2 cache 1024K; L3 cache 33792K
    \item 96 GB of main memory 1.880 GB/core, 12x 8GB 2667Mhz DIMM (216 nodes high memory, 10368 cores with 7.928 GB/core)
    \item 100 Gbit/s Intel Omni-Path HFI Silicon 100 Series PCI-E adapter
    \item 10 Gbit Ethernet
\end{itemize}

The processors support well-known vectorization instructions such as SSE, AVX up to AVX–512.

\subsection{PhysiCell} 
PhysiCell \cite{physicell_2018} is an open-source multiscale multicellular simulation framework written in C++ with minimal dependencies.
It was developed as an agent-based modeller where the agents are cells with properties like secretion/uptake values for different substrates, mechanical properties, growth rates, who interact with the tissue environment and with the other cells.
For this project, we are using PhysiCell version 1.6.0 and its source code is available at: \href{https://github.com/MathCancer/PhysiCell/releases/tag/1.6.0}{https://github.com/MathCancer/PhysiCell/releases/tag/1.6.0}.

PhysiCell is parallelized using OpenMP~\cite{openmp} as a shared memory
parallelization model. Recent works also have ported its diffusion solver to MPI (Message passing interface)~\cite{mpi} to allow scaling to several compute nodes~\cite{Saxena_biofvmx_2021}. However, the complete refactoring of PhysiCell to MPI has not been completed and thus, in this work, we focus on analyzing and optimizing the OpenMP version of PhysiCell. 

We use GCC~8.1 and OpenMPI~3.1.1 running atop the SUSE Linux Enterprise Server 12 SP2 OS 
to simulate the \texttt{heterogeneity} sample project in a 3D setup that starts with over 3900 cells in a simulation box of 1500~{\textmu}m per side and a total 
simulated time of 1440~minutes for the identification of regions to be optimized, and 12 days for final optimization impact validation. 

\subsection{Methodology}

The work presented here follows the methodology defined in the Performance and Optimization Center of Excellence (POP CoE). 
This implies that a detailed performance analysis guides the optimizations. 
The performance analysis process starts by defining the focus of analysis (FoA), the region of interest that will be analyzed.
Then, at the FoA, we apply the POP efficiency metrics~\cite{pop_metrics}. 
The POP efficiency metrics are a set of performance metrics that identify to which extent different factors affect the performance. 
They are mutually exclusive, hierarchical, and multiplicative.

\begin{figure}[htbp]
\centerline{\includegraphics[width=\linewidth]{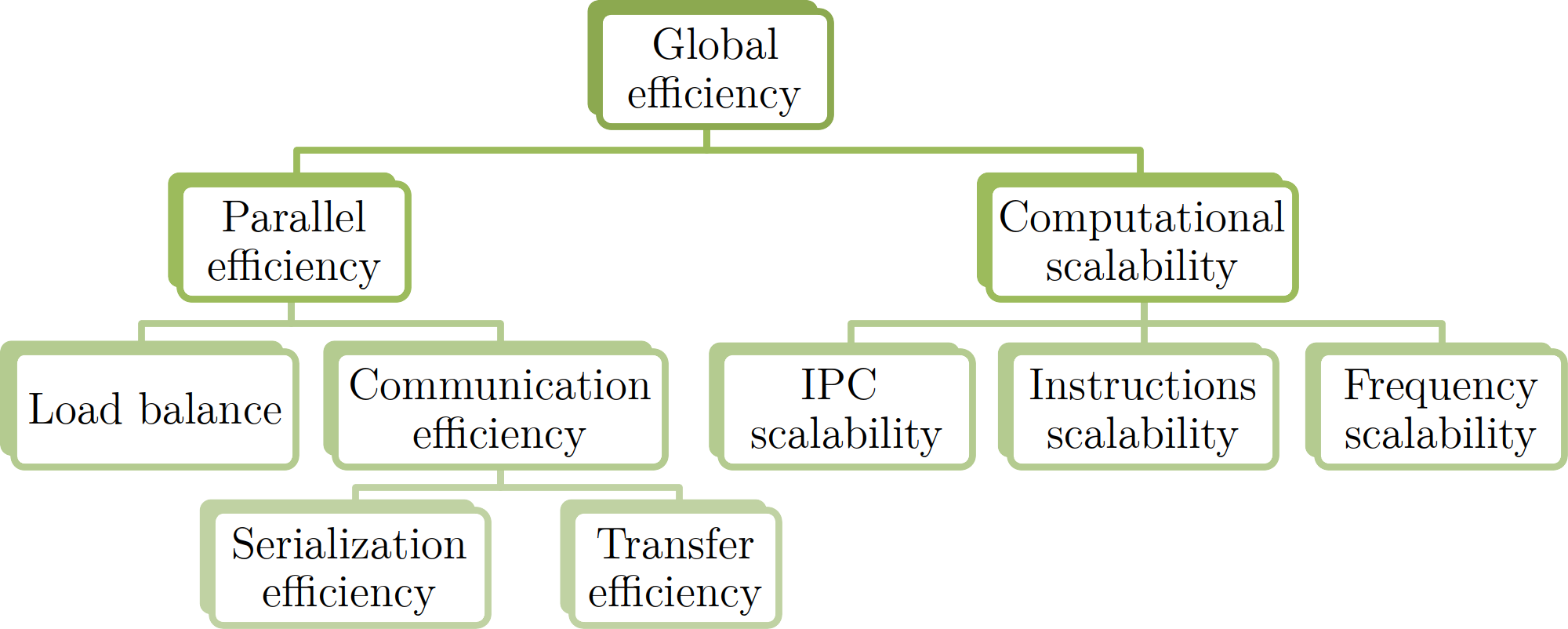}}
\caption{Hierarchy of POP Efficiency metrics.}
\label{fig:metrics}
\end{figure}

In Figure~\ref{fig:metrics} we can see the hierarchy of POP metrics. We can find a complete definition and how each metric is computed in the work of Garcia-Gasulla et al.~\cite{performance_CFD}. Below we give a short overview of the meaning of each metric when analyzing an OpenMP code. There are two kinds of efficiency metrics, efficiencies, and scalability; efficiencies are calculated based on the current execution, while scalabilities are computed based on a base case assuming an ideal computation scaling is expected. In the latter, for the base case, we presume scalability of 100\% and can work for strong and weak scaling.

\begin{itemize}
\item Parallel efficiency: This is the efficiency loss due to the overheads inherent to the parallelization.

\begin{itemize}
\item Load Balance: This is the efficiency loss due to an uneven distribution of workload among threads.
\item Communication efficiency: This is the efficiency loss due to the non-instantaneous nature of synchronization between threads.
\end{itemize}

\item Computation scalability is the efficiency lost in pure computation time relative to a base case. It is divided into three fundamental factors: instructions, IPC (Instructions per Cycle), and frequency.

\begin{itemize}
\item Instruction scalability: is the scalability of the number of instructions executed during the compute time with respect to the base case.
\item IPC scalability: is the scalability of the IPC achieved during the compute time with respect to the base case.
\item Frequency scalability: is the scalability of the frequency during the compute time with respect to the base case. 
\end{itemize}

\end{itemize}

A low value in one of these metrics will show the analyst what inefficiencies must be studied in detail to understand where and why they happen. We do this analysis based on execution traces obtained with Extrae~\cite{performance_opt} and analyze them using Paraver~\cite{paraver,paraver_web}. Once we have determined the main scalability issues or bottlenecks for that version, we discuss some recommendations with the developers. If agreed, the corresponding optimization is implemented and evaluated. After this, the performance analysis process can start over.

\section{Analysis and Optimization of Concurrent Memory Allocations}
\subsection{Focus of Analysis and Efficiency Metrics}

\begin{figure}[htbp]
\centerline{\includegraphics[width=\linewidth]{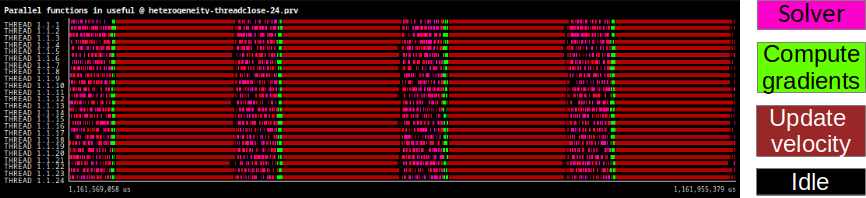}}
\caption{Focus of Analysis: Trace showing parallel functions for three iterations of PhysiCell.}
\label{fig:foa}
\end{figure}

Following the methodology, the first step is to determine the Focus of Analysis (FoA), in Figure~\ref{fig:foa} we can see a trace of three iterations of an execution of PhysiCell with 24 OpenMP threads. Each row represents one of the OpenMP threads and the X axis is the execution time. The different colors correspond to the different parallel functions.

The application presents an iterative pattern, for that reason we select the FoA
as one of the iterations. However, we still consider in our analysis the dynamic nature of the problem, because the number of cells can increase or decrease as the simulation advances.

Next step is to compute the efficiency metrics defined by the POP efficiency model. In table \ref{tab:efficiency-original} we show the efficiency metrics for the FoA of PhysiCell. Each column corresponds to one execution with a different number of OpenMP threads, the different rows are the different efficiency metrics.

\begin{table}[hbtp]
\caption{Efficiency metrics for original FoA of PhysiCell.}
    \label{tab:efficiency-original}
    \centering
    \includegraphics[width=1\linewidth]{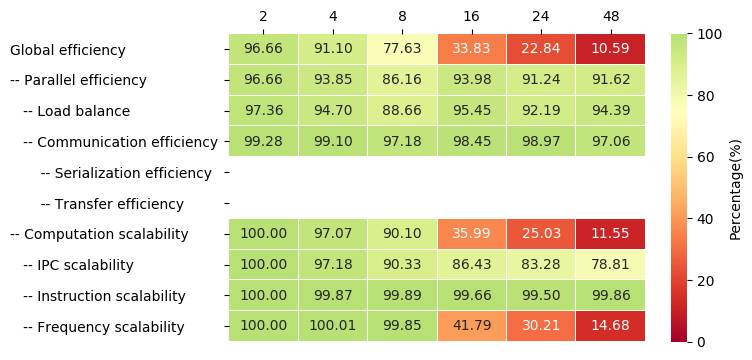}
\end{table}

We observe that the main problem limiting scalability is the drop of frequency, it decreases as the number of threads used increase. With 48 threads reaches a 14\% of the frequency measured with 2 threads. 

\begin{figure}[hbtp]
    \centering
    \includegraphics[width=\linewidth]{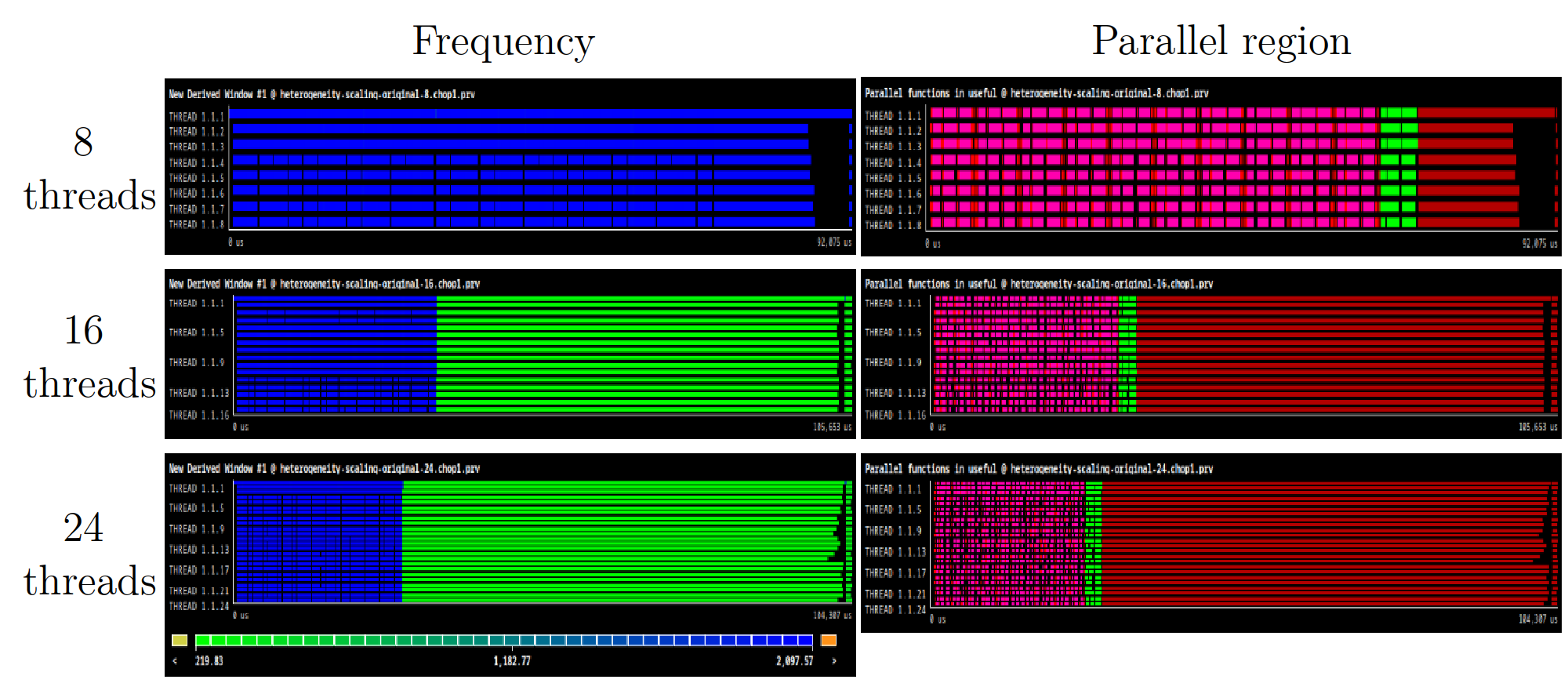}
    \caption{Traces comparing frequency values for executions with 8, 16, and 24 threads.}
    \label{fig:prv-freq-scalability}
\end{figure}

With a detailed analysis we try to determine where and why this decrease in the frequency happens.
Figure \ref{fig:prv-freq-scalability} shows traces visualizing the value of cycles/$\mu s$ (frequency) for 8, 16, and 24 threads executions. The top row corresponds to 8 threads; the middle row corresponds to 16 threads; the bottom row corresponds to 24 threads. On the left, visualization of frequency for each computation burst with a color gradient that goes from green for low values to blue for high values. On the right, the same traces showing parallel functions help relate regions on the left to their corresponding parallel function.

While we see the nominal frequency of MN4 in the trace of 8 threads, the traces for 16 and 24 threads clearly show a drop of the  cycles/$\mu s$ value in the update velocity region. 
    
The study of the traces leads to some observations: 1)The frequency drop is localized in one computation region; 2) The value of the frequency decreases as the number of resources increases; 3) The decrease of frequency significantly impacts the duration of the region, clearly being the bottleneck that prevents the code from scaling.

\subsection{Characterizing the cause of the frequency decrease}

To determine the cause of the decrease of frequency when increasing the number
of threads, we discard some of the common causes. It can not be preemption
(another process using the CPU where the thread is running)   because it is
localized in a code region and a preemption would be seen along the whole
execution. Checking the memory consumption we also discard that the code is
swapping data to disk because it would have exhausted the node memory.

Considering that the frequency decrease appears as we increase the number of threads running in the node, we can assume that the cause is related to a shared resource inside the node that is being stressed.

Analyzing the region of code suffering from the frequency drop and knowing that it should be related to a shared resource, we determine that the performance drop is related to the high number of small memory allocations and deallocations executed.

PhysiCell uses the overloading operator feature of C++ to redefine operations on variables of type \texttt{vector<double>}. Listing~\ref{lst:operator-overload} shows an example of overloading the operator that computes an addition between two vectors. 

\begin{lstlisting}[language=c++, caption={Overloading implementation of the operator + in PhysiCell}, label={lst:operator-overload}]
std::vector<double> operator+( const std::vector<double>& v1 , const std::vector<double>& v2 )
{
 std::vector<double> v = v1;
 for( unsigned int i=0; i < v1.size() ; i++ )
 { v[i] += v2[i]; }
 return v; 
} 
\end{lstlisting}

\begin{lstlisting}[language=c++, caption={Extract of function
\texttt{is\_neighbor\_voxel(...)}, creation of corner point variable using overloaded operators.}, label={lst:vector-creation-operation}] 
std::vector<double> corner_point = 
  0.5*(my_voxel_center+other_voxel_center); 
\end{lstlisting}

In Listing~\ref{lst:vector-creation-operation} we show a code where the overloaded operators are used. In a single line of code a variable of type \texttt{vector<double>} is created, and it is assigned the result of the product of a scalar by the addition of two vectors.

If we analyze in detail the code in Listing~\ref{lst:operator-overload}, we can see that for every invocation of the overloaded operator + a memory allocation is done in line 3 and the corresponding deallocation in line 6. 
This means that the single line of code of Listing~\ref{lst:vector-creation-operation} causes three allocations.

As we said, our problem works in a 3D environment, and vectors of three positions are used to place elements in the 3D space. Therefore, a very high number of operations are performed on small vectors. 

\subsection{Conclusions of the analysis}
Our study indicates that a high number of concurrent memory allocations and releases in the code are responsible for the lack of scalability of the code. These memory allocations are produced by the overloaded operators that operate on small vectors.
To overcome this bottleneck, we recommend the developers implement one of the following optimizations:
\begin{enumerate}
    \item Use standard library operators to manage vector positions
    instead of the overloaded ones;
    \item Reimplement the overloading of operators for \texttt{vector<double>}, so no extra memory is allocated to perform them. 
    \item Use an alternative library for dynamic memory management that manages concurrency better.
\end{enumerate}

The solutions preferred for an optimized version of PhysiCell are 1 or 2. These solutions, however, require a major refactor of the code. To prove the benefits of optimizing PhysiCell and encourage the developers to implement it, in the following section we apply solution number 3 and evaluate the results. 

\subsection{Optimization}

The library chosen for the proof of concept is jemalloc a "general purpose malloc(3) implementation that emphasizes fragmentation avoidance and scalable concurrency support" \cite{jemalloc-website}. This library is a good candidate for the proof of concept, as it promises to handle concurrency better and is easy to integrate with PhysiCell, as a preloaded library. Using jemalloc with PhysiCell does not require recompilation or code changes.

The execution of PhysiCell with Jemalloc requires adding the path of the library
to the \texttt{LD\_PRELOAD} environment variable before executing the program.

\begin{figure}[hbtp]
    \centering
    \includegraphics[width=\linewidth]{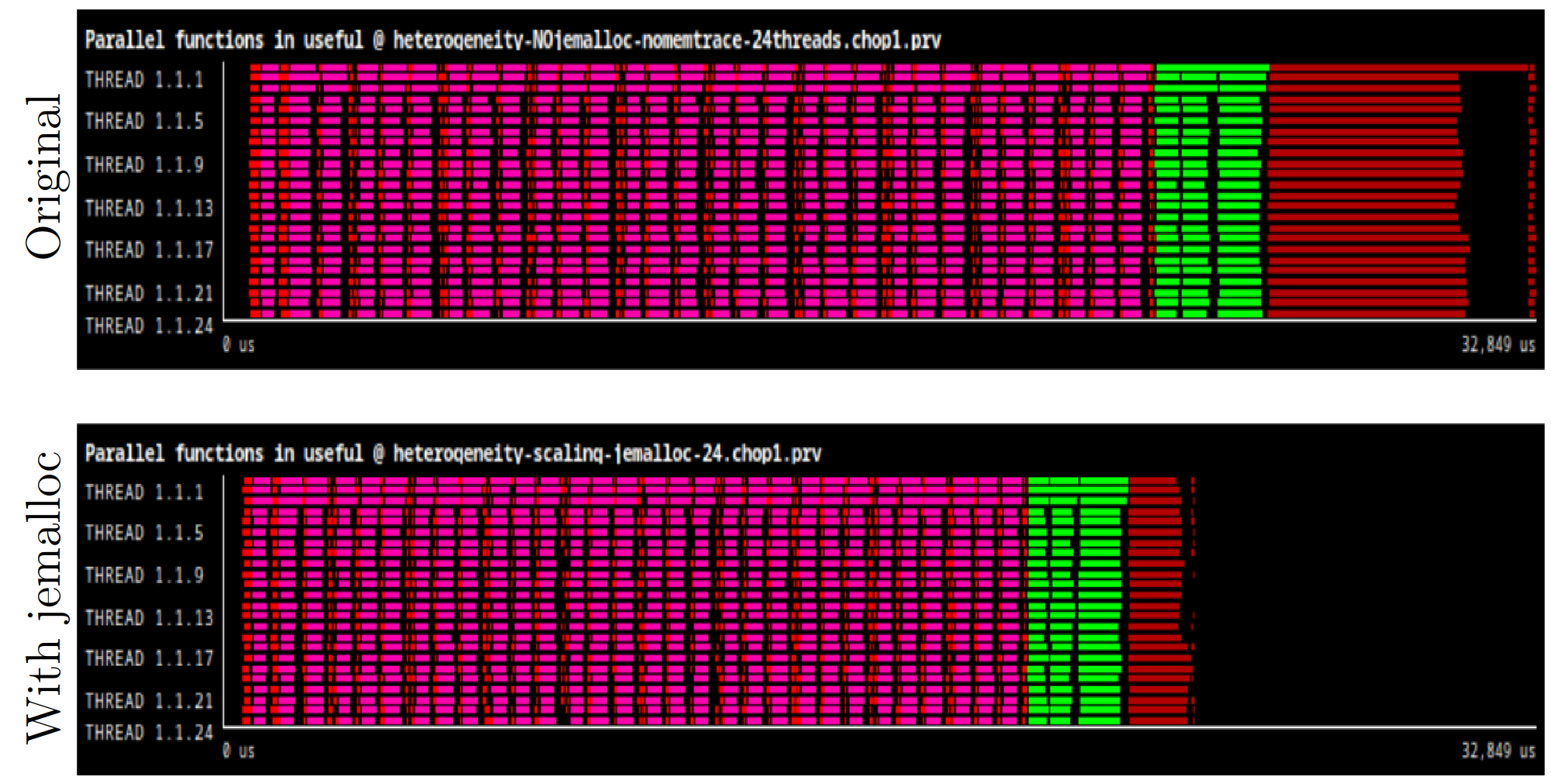}
    \caption[Execution trace of FoA comparing original and Jemalloc version.]{\textbf{Execution trace of FoA comparing original and Jemalloc version.}  Top trace corresponds to original execution, bottom trace corresponds to execution with jemalloc.  Colors indicate parallel function.}
    \label{fig:prv-compare-jemalloc}
\end{figure}

To evaluate the optimized version we execute the same use case preloading the jemalloc library. Figure \ref{fig:prv-compare-jemalloc} compares
execution traces of the FoA of both versions (top trace original execution,
bottom trace execution with jemalloc). The Focus of Analysis performs 30\%
faster than the original one in the case of 24 threads. The update velocity region (big burst in dark
red) reduces its execution time drastically.  The solver region (pink bursts)
also shows a reduction of the execution time, which indicates that the problem
of the memory allocations actually impacted the whole code.

\begin{figure}[hbtp]
    \centering
    \includegraphics[width=0.85\linewidth]{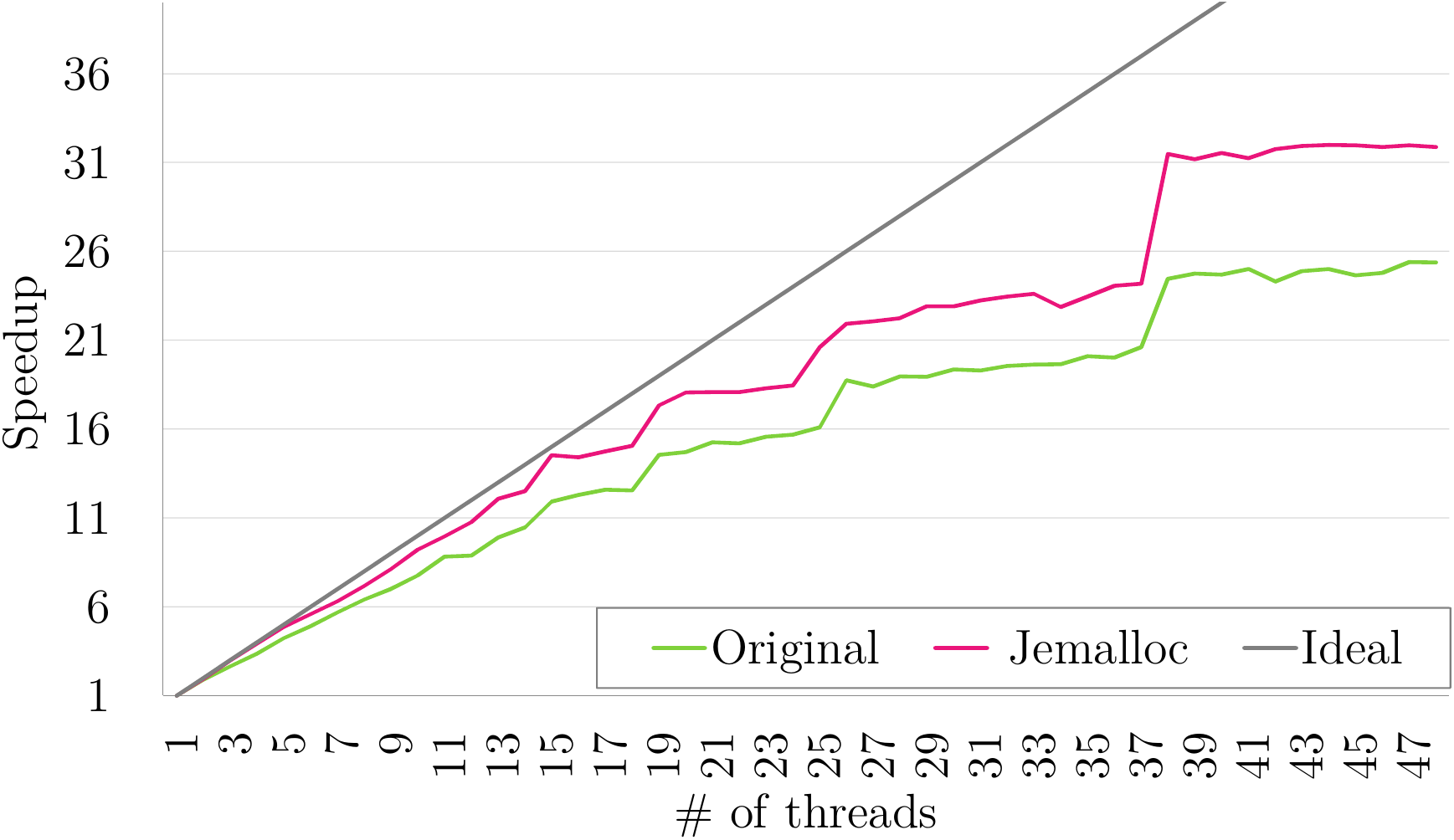}
    \caption{Strong scalability plot for 1-48 threads, original and jemalloc
    version.}
    \label{fig:plot-scalability-originalandjemalloc}
\end{figure}

Scalability is also improved, as expected.  The plot in Figure~\ref{fig:plot-scalability-originalandjemalloc} shows the strong scalability for
original and Jemalloc versions. In the X axis we can see the number of threads of the execution from 1 to 48; in the Y axis we can see the speedup with respect to the 1-thread execution of each series.  The version with jemalloc shows
speedups closer to the ideal one.  

\begin{figure}[hbtp]
    \centering
    \includegraphics[width=0.85\linewidth]{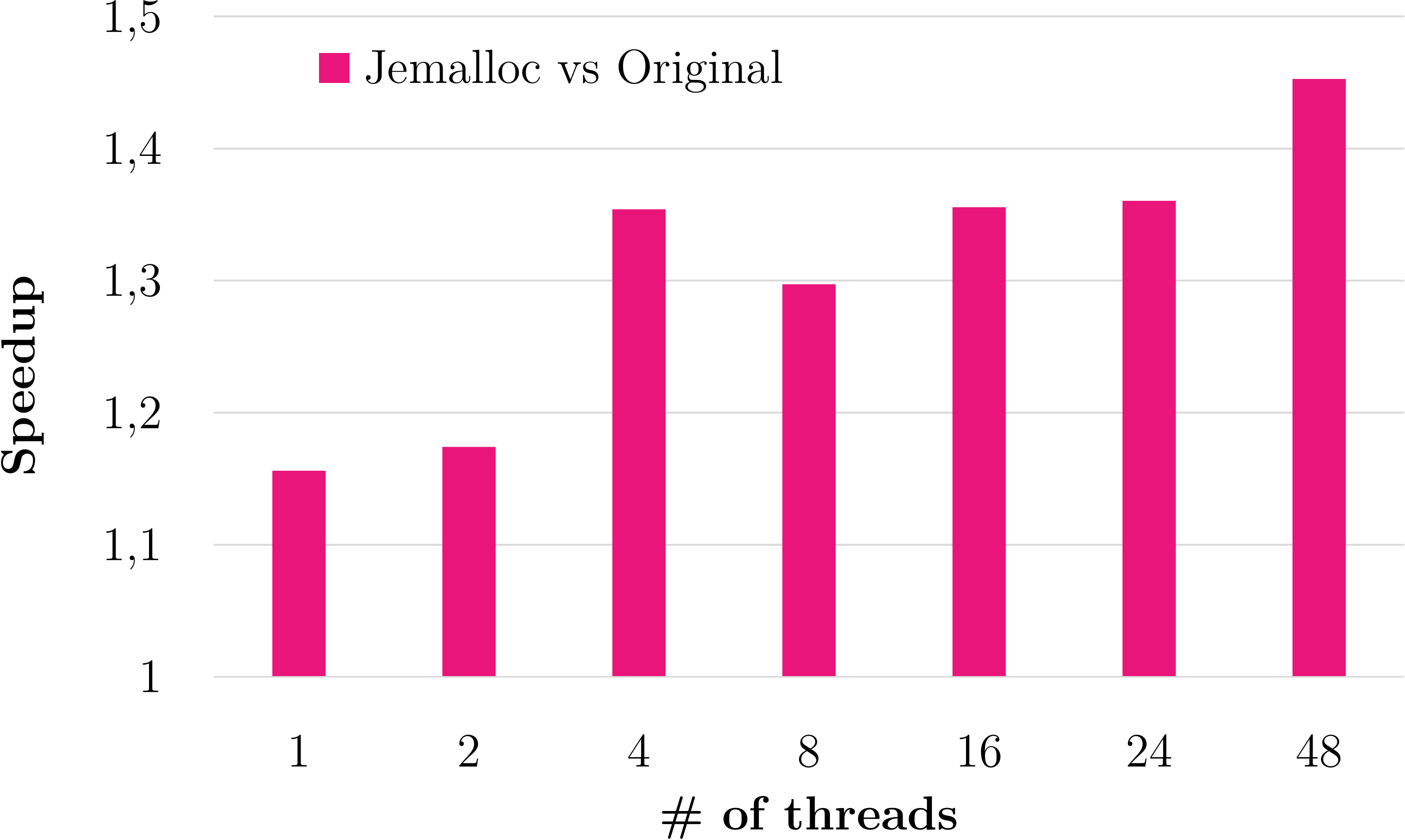}
    \caption{Speedup of jemalloc version respect original.}
    \label{fig:plot-speedup-bw-original-jemalloc}
\end{figure}

In Figure \ref{fig:plot-speedup-bw-original-jemalloc} we show the speedup
of the jemalloc execution with respect to the original one. We see a speedup of 1.36x (reduction of runtime by 27\%) for
executions with 24 threads or 1.45x (reduction of runtime by 31\%) for
executions with 48 threads.

These results indicate that the optimization resolves the problem limiting
scalability and that the factor limiting the scalability is the concurrency of frequent memory allocations.

\section{Analysis and Optimization of Load Imbalance}
\subsection{Detection of the problem}

We now start a new iteration of the performance analysis and optimization loop. We obtain the efficiency metrics of the same execution using the jemalloc library (table~\ref{tab:efficiency-jemalloc}). The first thing we observe in this table is that we no longer see the frequency issue that appeared in the previous table. In this case, we see that the lowest efficiency value is the load balance metric.

\begin{table}[hbtp]
    \caption{Efficiency metrics for FoA of PhysiCell with Jemalloc.}
    \label{tab:efficiency-jemalloc}
    \centering
    \includegraphics[width=\linewidth]{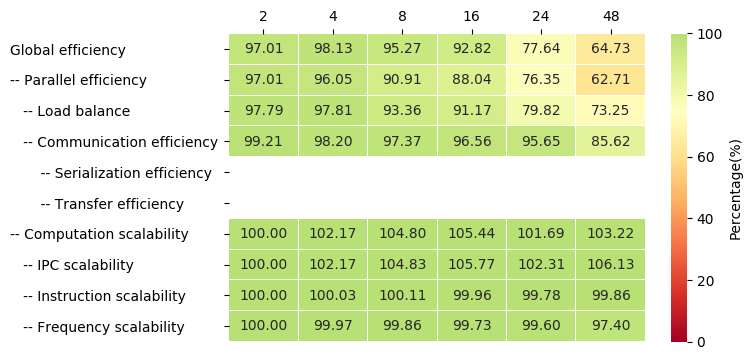}
\end{table}

\begin{figure}[hbtp]
    \centering
    \includegraphics[width=\linewidth]{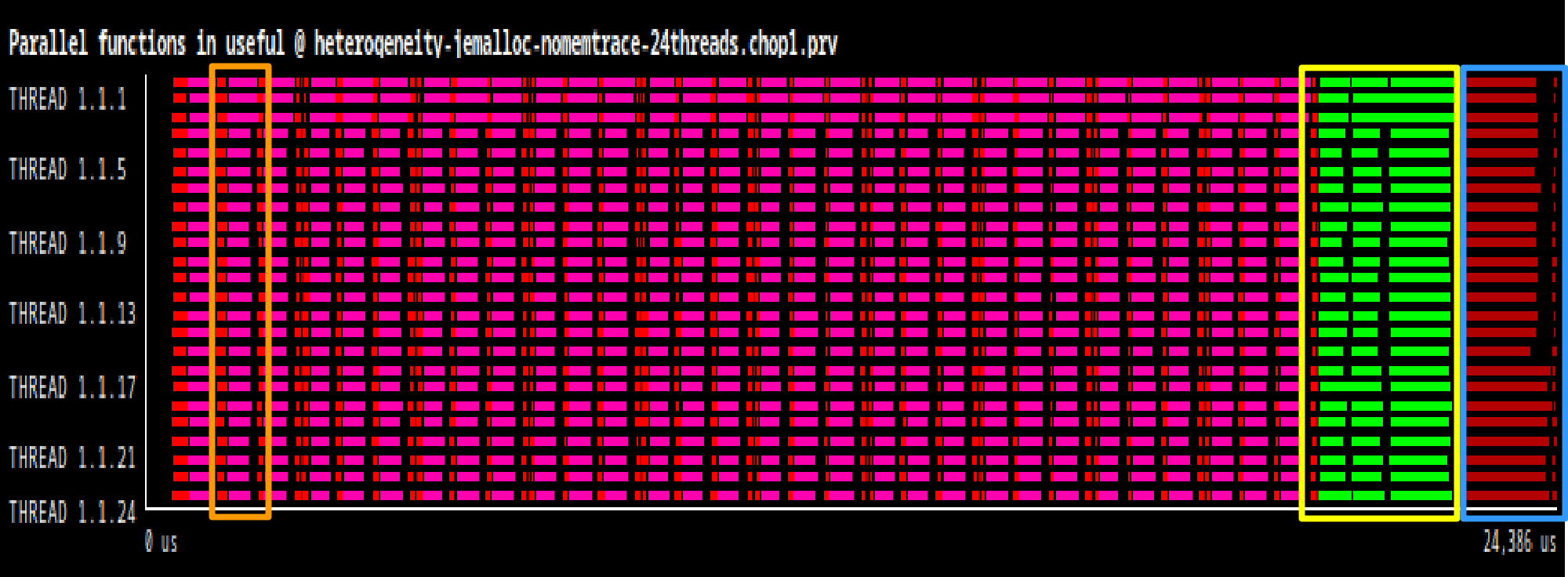}
    \caption[Paraver trace from PhysiCell with jemalloc execution, with regions
    of imbalance marked.]{Paraver trace from PhysiCell with jemalloc
    execution, with regions of imbalance marked.  Visualization is parallel
    functions. Own elaboration.} 
    \label{fig:prv-li-region-id}
\end{figure}

To provide suggestions on how to solve the load balance issue, first, we need to know its source. To do this, we analyze the trace of the FoA of the new execution in detail. In Figure~\ref{fig:prv-li-region-id} we show the parallel regions of one iteration of PhysiCell. 
We detected three different regions with load imbalance and identified them with color squares in the image.
Orange square (left) shows that first threads always take longer to complete. It corresponds to the microenvironment solver function. Yellow square (middle) shows that first threads always take longer to complete. It corresponds to the compute gradients function. Blue square (right) marks that the last threads show a longer execution time.  It corresponds to the cell's update velocity function.  

\begin{figure}[hbtp]
    \centering
    \includegraphics[width=\linewidth]{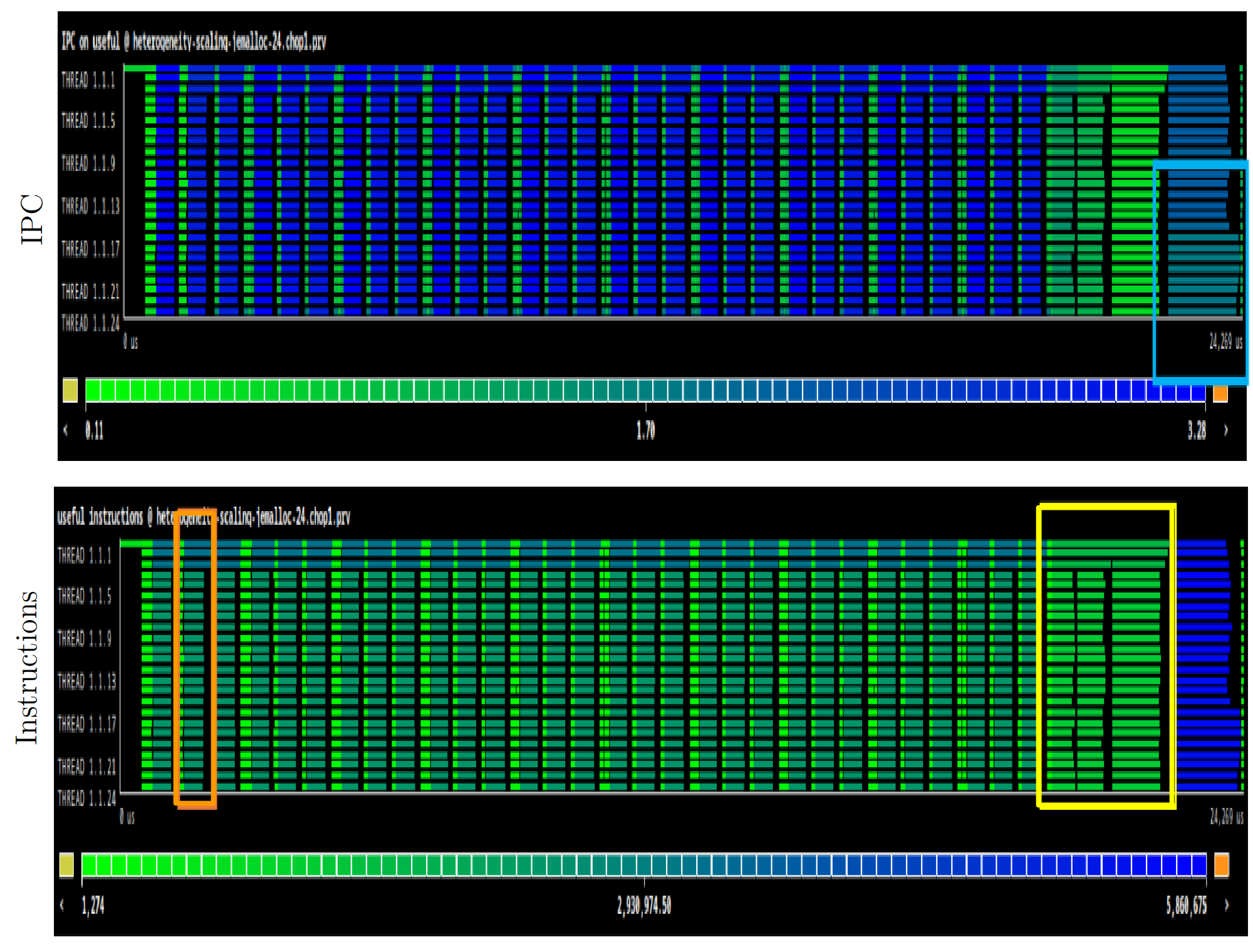}
    \caption{Paraver traces showing IPC (top) and number of instructions (bottom) per thread. } 
    \label{fig:prv-inst-imbalanceregion}
\end{figure}

When load imbalance occurs, it can have two explanations: either some threads have more work to do (i.e., they execute more instructions), or they compute the same work slower (i.e., at a lower IPC). To understand in which case we are looking at, we measure the number of instructions executed by each thread in each parallel region and the IPC obtained. We show these in the traces depicted in Figure~\ref{fig:prv-inst-imbalanceregion}.
The bottom trace shows a color gradient for the number of instructions, where green means lower and blue means higher. We can observe that for regions yellow and orange the first three threads execute more instructions than the other threads. 

In the top trace of Figure~\ref{fig:prv-inst-imbalanceregion} we show the IPC
also in a gradient code color. We observe that while the yellow and orange
square regions show the same IPC for all threads, at the blue square region the trace shows a clear
difference in the IPC of the threads that take longer to execute.  The IPC is
lower, meaning that the same number of instructions needs more time to be executed compared to
another thread.

Based on this we can conclude that for the regions orange and yellow, the source
of the load imbalance is the computational load. On the other side, for the blue region the source of load imbalance is the IPC.

\subsection{Study of the load imbalance due to workload}
When using \texttt{parallel for} pragmas to parallelize OpenMP code and deal with load imbalances, the common approach is to use a different loop scheduler that helps balance the workload among the threads.

\begin{lstlisting}[language=c++, caption={Parallelization of microenvironment solver}, label={lst:microenvironment}]
# pragma omp parallel for
for(unsigned int k=0; k<M.z_coord.size(); k++)
    {
    for(unsigned int j=0; j<M.y_coord.size(); j++)
    {
        ...
        \\Some work
        ...
    }
}
\end{lstlisting}

But once we identified the parallel loop displaying the load imbalance, we see that it has the structure shown in Listing~\ref{lst:microenvironment}. This code snippet shows that the solver process is parallelized on the outermost loop of the domain traversal. Knowing that our specific use case contains a domain of $75 \times 75 \times 75$ voxels. There are only 75 chunks of workload to distribute among at most 48 threads.

Our conclusion is that the load imbalance cause is the work's granularity being too coarse. This behavior happens
both for the solver phase and the compute gradients function, as the parallelization strategy is the same.

Our suggestion is to implement a finer parallelization to better
distribute workload between threads and avoid load imbalance, independently of
the number of cores that the machine has.

\subsection{Optimization of the load imbalance due to workload}

Looking at the original code, we see that we can still parallelize inside the
loop.  There is a nested loop right after the outer loop that iterates through another domain axis.  Our implementation consists of collapsing the two outer loops, in order to increase the workload from \texttt{M.z\_coord.size()} to \texttt{M.z\_coord.size() * M.y\_coord.size()}.  
With this, we aim to reduce the grain size, and a higher number of finer chunks will be distributed, achieving a more balanced workload.

OpenMP specification provides the \texttt{collapse} clause, allowing to parallelize multiple nested loops without introducing nested parallelism.  Our implementation applies this clause to the parallel regions affected by the load imbalance.

\begin{figure}[hbtp]
    \centering
    \includegraphics[width=\linewidth]{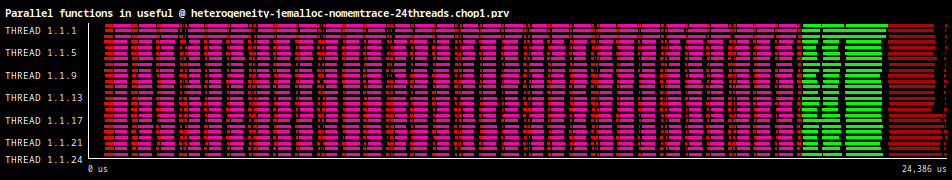}\\[0.1cm]
    \includegraphics[width=\linewidth]{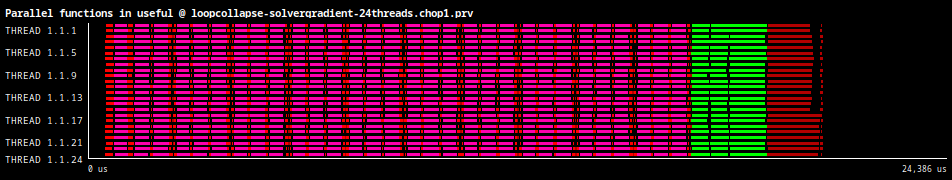}
    \caption{Trace comparison between original + jemalloc (top) and original + jemalloc + collapse (bottom).}
    \label{fig:prv-optimization1-comparison}
\end{figure}

In Figure~\ref{fig:prv-optimization1-comparison} we compare the previous version with our optimized version using the collapse clause. In the top trace we see one iteration of the original version with the jemalloc library, and in the bottom trace we see the same iteration using the jemalloc library and with the \texttt{collapse} clause. We can see how the microenvironment solver (pink) and the compute gradients function (green) are better balanced and therefore also faster to compute in the new version.

\subsection{Study and optimization of the load imbalance due to IPC}

The region affected by the load imbalance due to IPC corresponds to the update
velocity parallel function. This code consists of a loop over all the cells to update its velocity based on their neighbouring cells. The parallelization strategy consists of a \texttt{parallel loop} distributing the computation of the total cells between the threads. 
There is no reason to think that the computation of the velocity of one cell is different from another. We need to find out why some cells negatively impact IPC when computing its velocity.

\begin{figure}[hbtp]
    \centering
    \includegraphics[width=\linewidth]{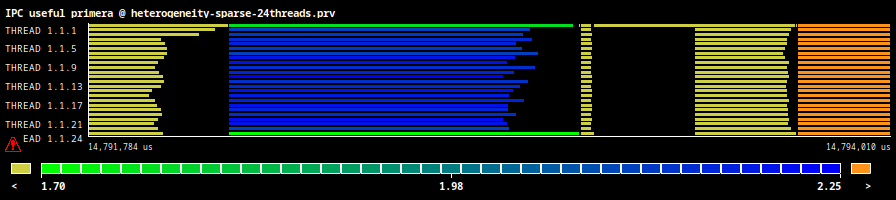} \\[0.1cm]
    \includegraphics[width=\linewidth]{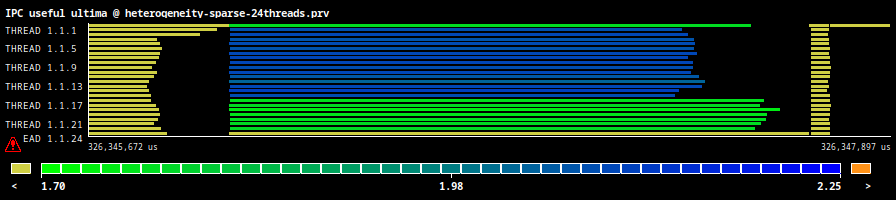}
    \caption[Execution traces, IPC visualization, zoomed-in update velocity
    region.]{Execution traces, IPC visualization, zoomed-in update
    velocity region. The top trace corresponds to the first iterations, bottom
    trace corresponds to an advanced point of the simulation.}
    \label{fig:prv-compare-ipc-iterations}
\end{figure}

In Figure~\ref{fig:prv-compare-ipc-iterations} we show the IPC at the update velocity region and we compare the same region at two different moments of the simulation.
These traces show a gradient for the value of IPC, for green meaning low values, and blue meaning high values. Yellow and orange represent lower out of range and upper out of range, respectively.
The top trace, that corresponds to one of the first iterations, shows only the
first and last thread with a lower IPC than the others. In contrast, the bottom
trace, that corresponds to an iteration of an advanced simulation point, shows more
threads with a lower IPC, starting from the last one and going backward.  

Another important observation is that the threads that are not
affected by the significant drop of IPC still show a slight decrease in their
IPC in the trace corresponding to the advanced point of the simulation.

To sum up:
\begin{enumerate}
    \item The number of threads that show a worse IPC increases as the
    simulation advances.  This effect spreads from the last thread and to the
    others backward.
    \item The other threads also show a slight decrease in IPC as the simulation
    advances.
\end{enumerate}

\begin{figure}[hbtp]
    \centering
    \includegraphics[width=\linewidth]{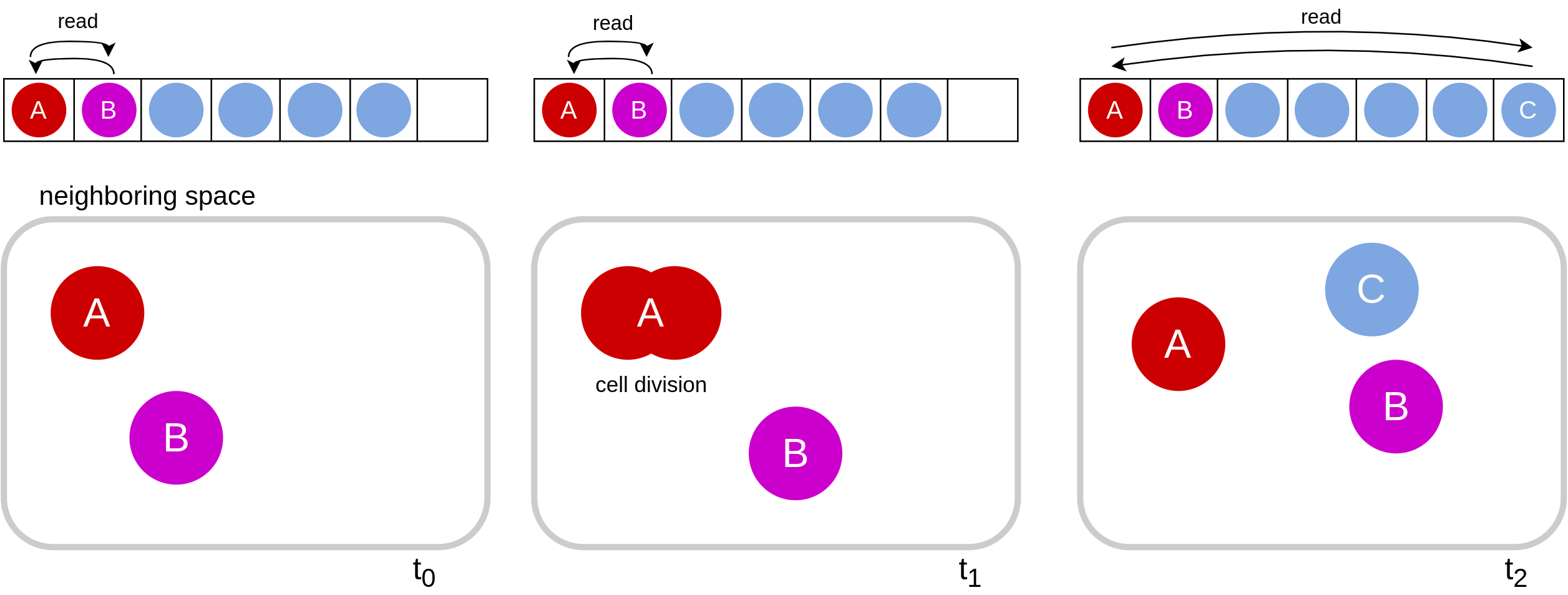}
    \caption{Representation of cell's position in the vector of cells and in a mesh slice at the initial
    simulation point.}
    \label{fig:cell-diagram-initial}
\end{figure}

Analyzing the data structures and the code we see that cells are stored in a vector, where the new created cells are appended at the end. This is represented in Figure~\ref{fig:cell-diagram-initial} where the new created cell C is neighbour of the cells A and B and is stored at the end of the vector. With the current parallelization that divides the vector of cells with a static schedule among the different threads, the runtime assigns to the last threads the cells created due to cell division during the simulation. 
As more cells divide, the proportional part of the vector that contains new cells grow, and
this is why the number of threads that compute these cells increases.  These
cells are more costly to compute because of their position in memory.  Finally,
the threads that do not have new cells are still affected because they need to
access information from the new cells, as they are neighbors of the original
ones.

For each new cell, PhysiCell adds a pointer to its data structure to the end of
the vector of cells.  Although these pointers are contiguous in the vector and
the cells they point to are computed by the same thread, they are physically
distributed throughout the 3-D mesh.  Therefore the physical neighbors are not
contiguous in memory.  This behavior makes accessing the neighbors of a cell to
compute its velocity very costly, as the program can exploit neither the spatial
nor the temporal locality, producing a high number of cache misses.

At this point we conclude that the objective of the optimization is not to solve
the load imbalance in the compute velocity function. This in fact would be achievable
by using a dynamic schedule in the parallel loop over the cells. Instead, the challenge is to achieve a good IPC for all the threads during all the execution.

We know that in an initial state, all cells contiguous in the vector of cells are also
neighboring in the 3-D mesh.  New cells resulting from division get added to the
end of the vector. In Figure~\ref{fig:cell-diagram-initial} when the program computes the velocity of cell A reads data
from cell B because they are neighbors.  When it computes the velocity of cell B
it already has the data needed of cell A in cache, because it was computed right before.

\begin{figure}[hbtp]
    \centering
    \includegraphics[width=\linewidth]{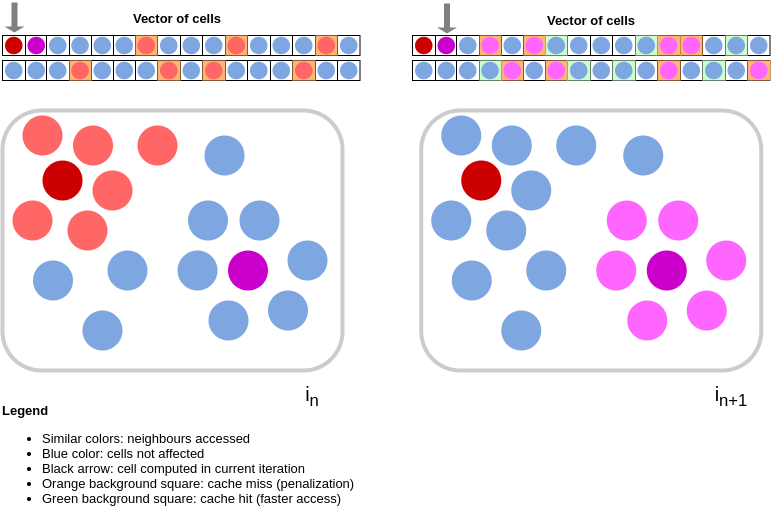}
    \caption[Representation of cell's position in the vector of cells and in a
    mesh slice at an advanced simulation point.]{Representation of
    cell's position in the vector of cells and in a mesh slice at an advanced
    simulation point.}
    \label{fig:cell-diagram-advanced}
\end{figure}

However, now let us jump to a more advanced simulation state, where
many new cells have been added to the simulation.  We see in Figure
\ref{fig:cell-diagram-advanced} two steps of the computation, performed by the
same thread, each corresponding to the computation of a different cell.  At
$i_n$ the cell marked with the arrow is computed, accessing data from all its
neighboring cells, painted in a lighter red.  This data is stored in the cache.
The same thread computes the next cell in the vector at the next iteration
$i_{n+1}$, marked with the arrow.  This cell now has neighbors that are entirely
different from the previous cell.  Therefore, accessing their data instead of
accessing the data already cached produces cache misses and is more costly.

Our suggestion to improve the IPC in the whole execution would be to store the cells in memory close to the cells they neighbour in the 3D environment, but this is not a trivial task and requires a considerable refactoring of the code.

For this reason we try to address the problem from the parallelization point of view.

\begin{lstlisting}[language=c++, caption={Parallelization of update velocities based on voxels}, label={lst:parallel_voxels}]
# pragma omp parallel for schedule ( dynamic , GS )
for (int voxelID=0; voxelID<agent.size(); voxelID++)
{
    //Check if voxel is empty
    if(agent[voxelID].size()>0)
    {
        //Compute voxel's cells
    }
}
\end{lstlisting}

As we already have a data structure that represents the 3D space, the voxels, we propose to parallelize the workload using voxels instead of the vector of cells. 
Then cells will be grouped and computed by
voxels, ensuring that neighboring cells are computed contiguously by the same
thread.  This parallelization strategy allows the threads to exploit the
temporal locality of the cell's data, producing more cache hits and faster
accesses.

\begin{figure}[htbp]
\centerline{\includegraphics[width=0.9\linewidth]{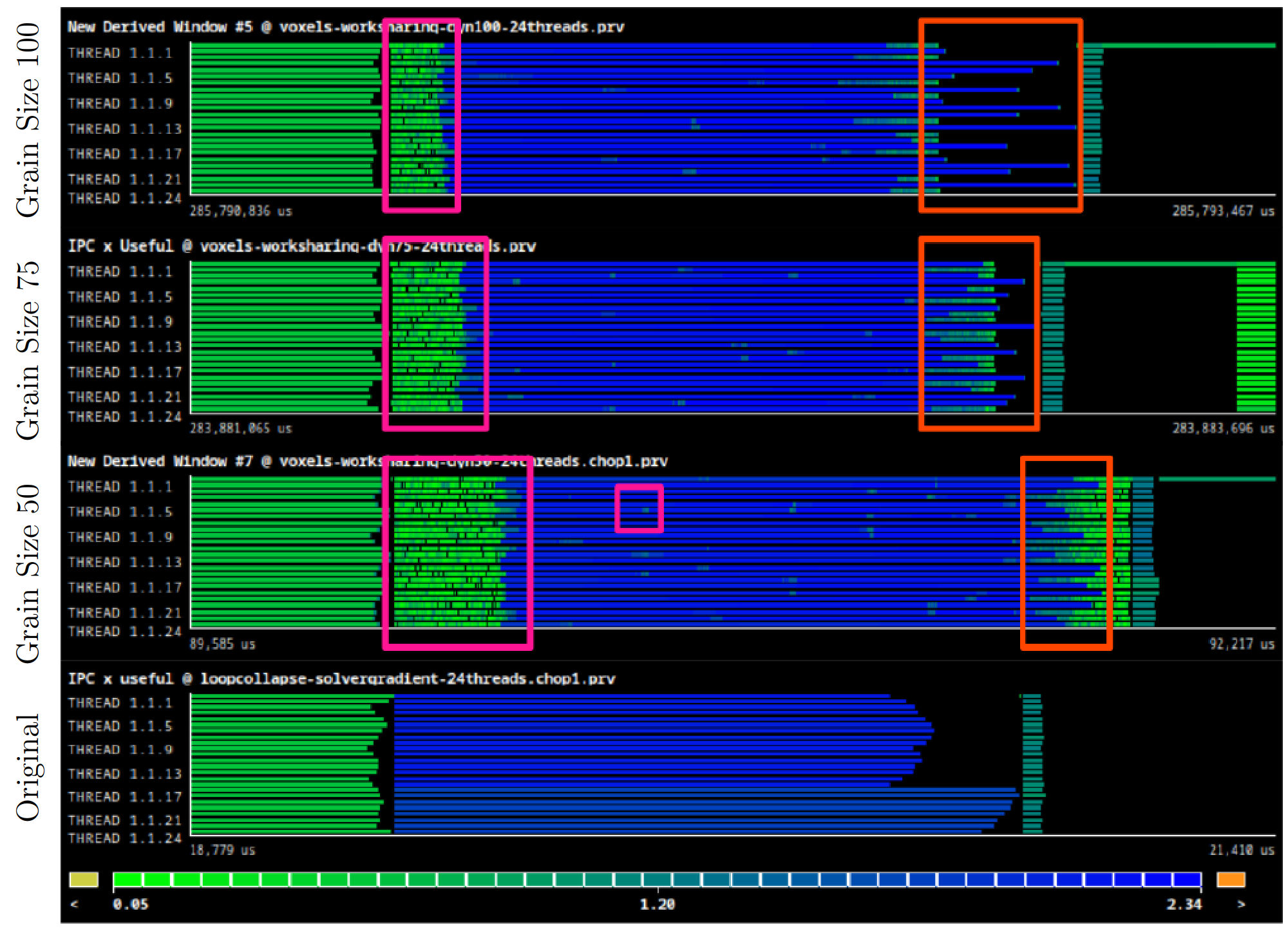}}
\caption{Trace comparison of a voxels based parallelization in update velocity function}
\label{fig:parallel_voxels}
\end{figure}

In Listing~\ref{lst:parallel_voxels} we show the parallelization proposed based in the voxels and we study the impact of the grain size (GS) in the performance. In Figure~\ref{fig:parallel_voxels} we show the traces obtained with different grain sizes compared with the original code. We can see that the new implementation has a worse performance than the original one. This is due to the computation of empty voxels at the beginning (pink squares) and at the end. Also, when using bigger grain sizes the impact of the empty voxels is lower but in this case the load imbalance at the end is what limits the performance.

\begin{figure}[htbp]
\centerline{\includegraphics[width=0.9\linewidth]{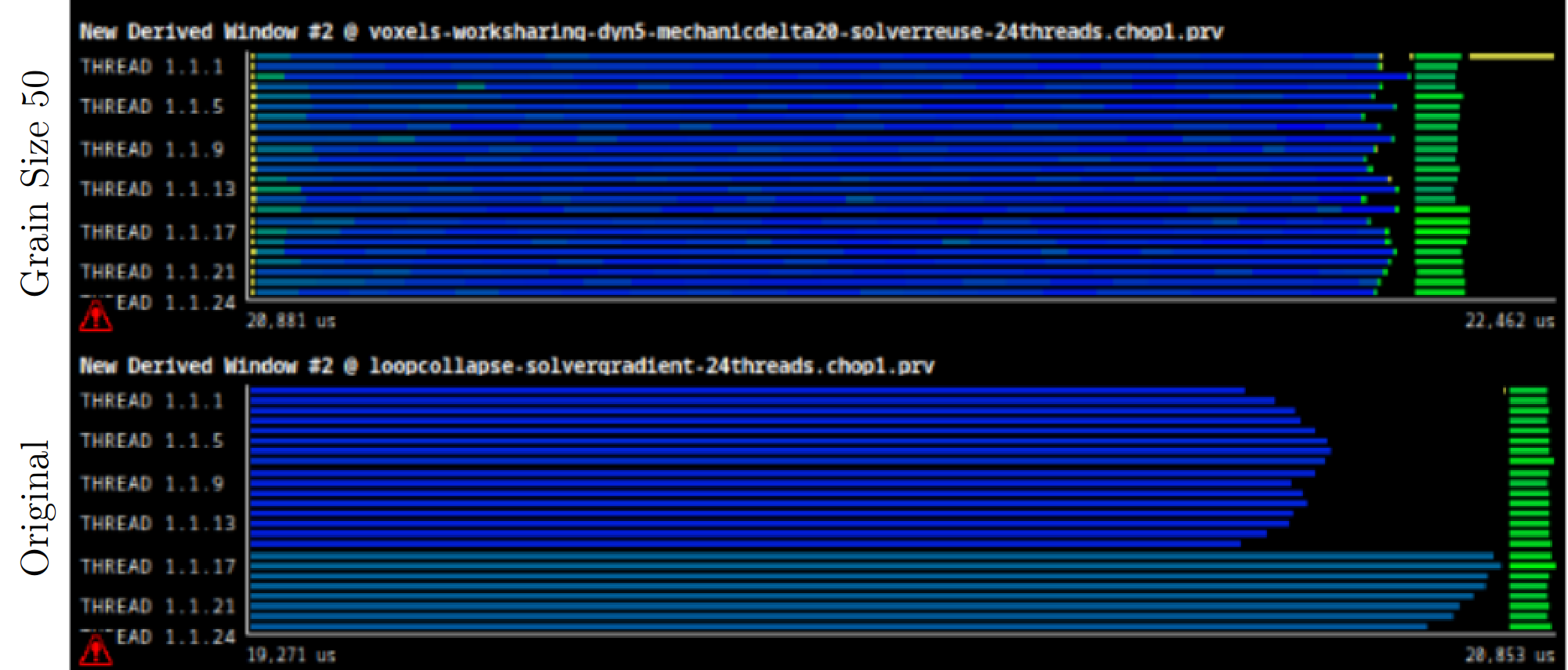}}
\caption{Trace comparison of non-empty voxels based parallelization in update velocity function.}
\label{fig:parallel_voxels2}
\end{figure}

In order to solve this issue we implement a new version that includes a list of non-empty voxels. This list of not empty voxels is updated during one of the solver processes, which already iterates through the 3-D mesh. In Figure~\ref{fig:parallel_voxels2} we show the trace obtained with this version we observe a small gain when using 24 OpenMP threads with respect to the previous version. However, due to the code added to compute the list of non-empty voxels the gain is not relevant in the overall execution.

This last optimization has not achieved relevant speedups in smaller experiments, but does present advantages when the simulations are longer and more cells appear. We have learned
some lessons: 1) Voxels (and, consequently, cells) should be computed in the
same order as found in the physical mesh. 2) A coarser grain exploits the
temporal locality much better; 3) The computation of empty voxels when working
through agent-based processes is useless and adds overhead, thus, should be avoided.

\subsection{Evaluation of load balance optimizations}

\begin{figure}[htbp]
\centerline{\includegraphics[width=\linewidth]{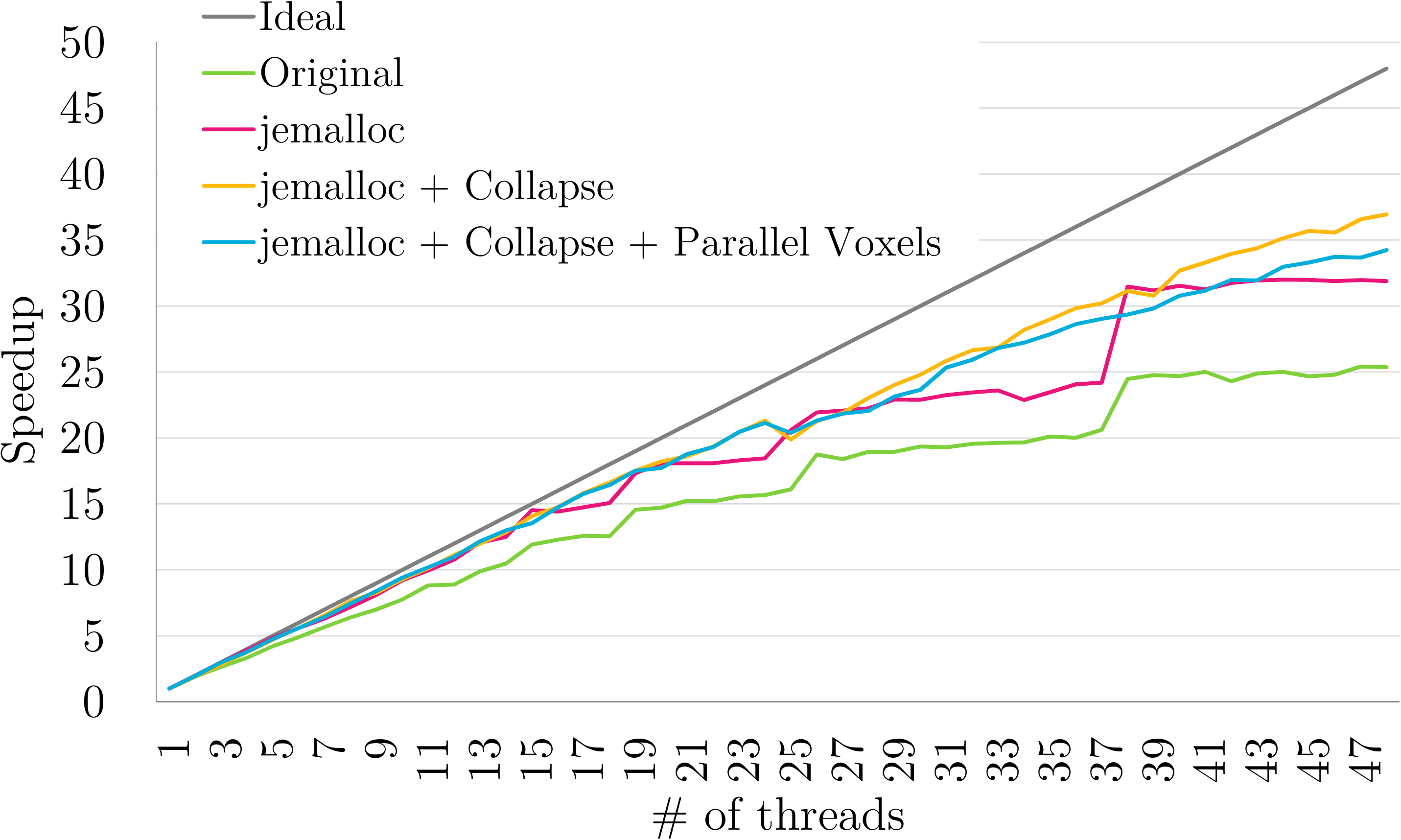}}
\caption{Scalability obtained by different versions up to 48 threads.}
\label{fig:scalability-final}
\end{figure}

In Figure~\ref{fig:scalability-final} we show the scalability of the different versions that we have shown in this work. In the $y$ axis is shown the speedup with respect to the execution with one thread of the same version. In the $x$ axis is shown the number of OpenMP threads used. Each speedup point is computed from 5 executions with different initial random seeds. Statistical variability of the measurements is below 5\%, so we decided not to include error bars.

We can see that for the versions including the \texttt{collapse} clause, the ``stairs'' that are shown in the original and jemalloc versions disappear. With the finer granularity parallelization, the efficiency is not tied to the number of resources used. The version that shows the best scalability includes the jemalloc and collapse optimizations, this version achieves a speedup of 37x with 48 OpenMP threads and 21x with 24 threads.



\begin{figure}[htbp]
\centerline{\includegraphics[width=\linewidth]{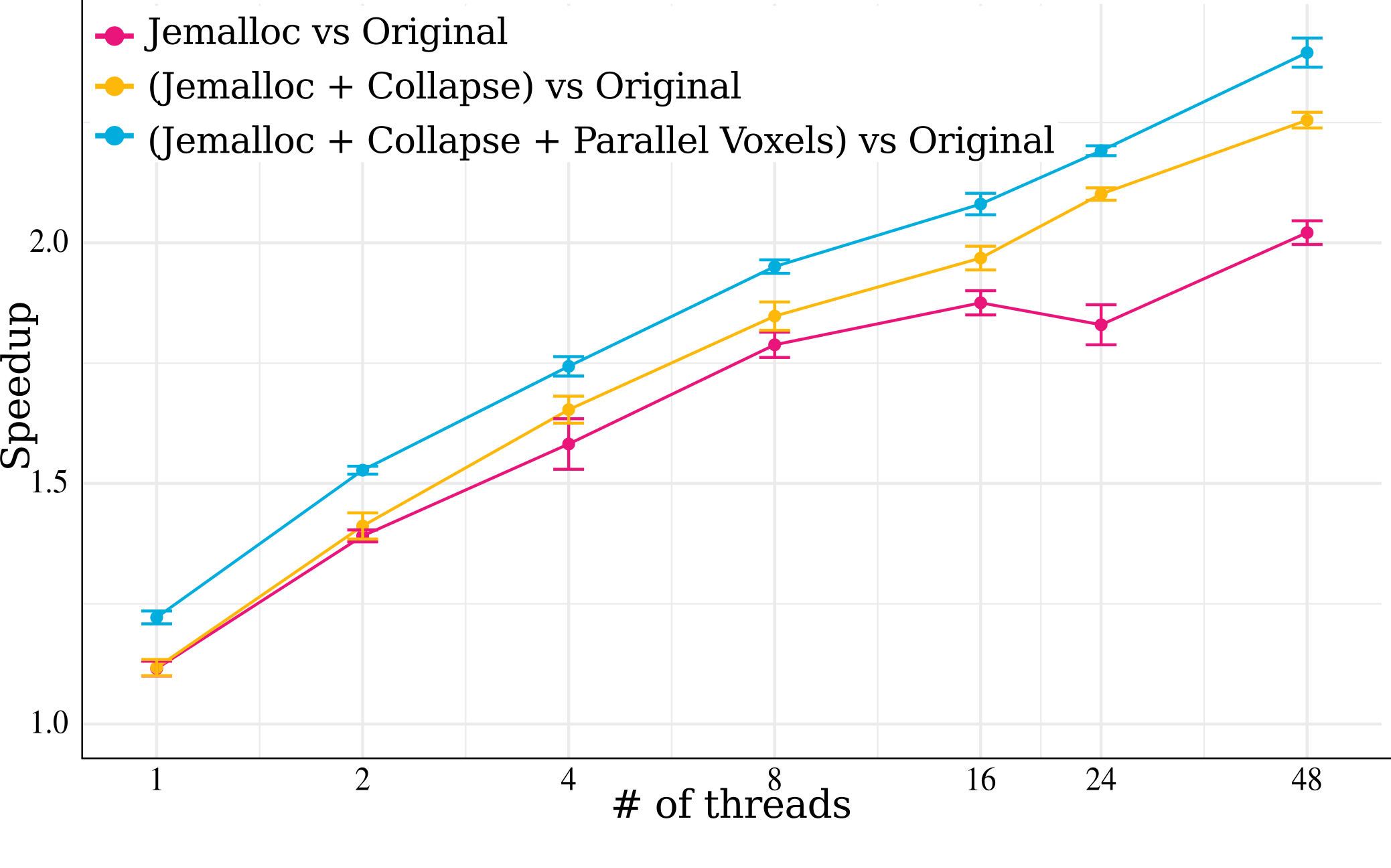}}
\caption{Speedup of optimizations wrt. original code for different thread configurations. Simulation of 12 days.}
\label{fig:speedup-final}
\end{figure}
Finally we simulate 12 days to reproduce a classical experiment of organoid growth\cite{realcase} to evaluate our optimizations in a production run.
In Figure~\ref{fig:speedup-final} we show the speedup of each version with respect to the original one. 
In the $y$ axis is shown the speedup, in the $x$ axis is shown the number of OpenMP threads used.  Each data point is the mean of 5 speedups and error bars are shown.
We can see how the optimized versions outperform the original code in all the
cases. 
With more threads used, the more relevant the optimization is, achieving a speedup of 2.4$x$ when using 48 threads combining the Jemalloc, Collapse and Parallel voxels optimizations and compared to the original version running with the same number of resources. 


\section{Conclusions and Lessons Learned}

Multiscale simulations represent one of the most computationally intensive fields of computational biology. In this context, relevant emerging paradigms, such as personalized medicine, are progressively demanding more sophisticated models and simulations, incorporating a large number of parameters to recapitulate the essential properties of biological systems appropriately \cite{Montagud_gigascale_2021}.

In the particular context of tissue modeling with PhysiCell, this type of expansion will involve the design of experiments that substantially increase the number of modeled cells (up to $10^9$), the number of substrates diffusing through the medium, and the complexity of molecular pathways to be quantified within each agent, inevitably leading to an exponential growth of computation required in each simulation.

These challenging scenarios will not only require the availability of larger
supercomputers, but also to use them efficiently. Therefore, improving the tools' performance without compromising their portability is crucial to use the HPC resources efficiently in those large-scale simulations.

This work presents three iterations of the performance analysis cycle leading to three optimizations. As our optimizations target quite common patterns in ABM codes, we share them as best practices to help developers implement more efficient tools.

\textbf{Memory allocation and deallocation contention.} We show that implementing custom operators that allocate and deallocate memory can compromise performance. If these operators are used very frequently combined with parallel execution, they generate contention in the memory allocating system. We encourage developers to avoid this practice. To demonstrate it without a major refactoring of the code, we show how using a memory managing library (jemalloc), we achieve a speedup of 1.45X when using 48 concurrent threads.

\textbf{Load imbalance, coarse granularity.} Load imbalance represents one of the main challenges that compromise the scalability of a given software. 
In the current scenario, we deal with a load imbalance produced by a coarse granularity of the parallel chunks of work. 
In this case, we advise against the common temptation of adjusting the workload partition to the given platform or hardware 
to avoid the loss of portability (i.e., to partition the work based on the number of computational resources). 
Instead, we propose to decrease the granularity of the parallel workload, for instance, by adding a collapse clause that 
allows parallelizing nested loops. With this change, combined with the previous one, we earn a speedup of 2.25x with respect to the original code.

\textbf{Load imbalance, heterogeneous IPC.} In the last analysis, we observe a load imbalance problem produced by the variation in the IPC achieved 
by the different threads as the simulation advances. Our focus, in this case, is not only solving the imbalance but also improving the memory 
locality that is producing the reduction in the IPC. After analyzing several implementations, we conclude that the best approach is to store 
cells in memory as close as possible to their position in the 3D environment, and at the same time to iterate when computing the cells as they 
physically appear in the 3D environment.  With our change, we get to a final speedup of 2.4x of all optimizations combined compared to original 
code (Figure~\ref{fig:speedup-final}).

We hope that the best practices presented here help developers better scale their ABM tools and other multiscale simulators to high-end computing nodes.

\section*{Acknowledgments}
This work has received funding from the EU Horizon 2020 POP2 (grant agreement No 824080), PerMedCoE (grant agreement No 951773), and by the Spanish Government (PID2019-107255GB).

\bibliographystyle{IEEEtran}
\bibliography{biblio}


\end{document}